\documentclass[11pt,twoside]{article}
\usepackage{asp2004}
\usepackage{psfig}
\usepackage{epsf}
\usepackage{graphics}
\usepackage{lscape}
\def\edcomment#1{\iffalse\marginpar{\raggedright\sl#1\/}\else\relax\fi}
\reversemarginpar

\begin{document}
\title{%Background Galaxy Detection with Large Telescopes\\ and Improved Optics\\
Total Opacity of Local Group Galaxies\\ and Large Scale Structure behind the Galactic Bulge}
\author{Rosa A.\ Gonz\'alez}
\affil{Centro de Radioastronom\' \i a y Astrof\' \i sica, UNAM, Morelia, Mexico}
\author{Benne Holwerda}
\affil{Kapteyn Institute, Groningen, The Netherlands}
\author{Laurent Loinard}
\affil{Centro de Radioastronom\' \i a y Astrof\' \i sica, UNAM, Morelia, Mexico}
\author{Ronald J.\ Allen}
\affil{Space Telescope Science Institute, Baltimore, MD, USA}
\author{S\'ebastien Muller}
\affil{Academia Sinica Institute of Astronomy and Astrophysics, Taipei, Taiwan}

\begin{abstract}
Recently, we have developed and calibrated 
the Synthetic Field Method to
derive total extinction through disk galaxies. The method is based on the
number counts and colors of distant background field galaxies that can be seen
through the foreground object. 
Here, we investigate how large (10-m) and very large (20 to 30-m), 
diffraction-limited, optical and infrared telescopes
in space would improve the detection of background galaxies behind Local
Group objects, including the Galactic bulge. 
We find that, besides and perhaps more important 
than telescope size,  
a well-behaved, well-characterized PSF would facilitate in general the
detection of faint objects in crowded fields, and greatly benefit several other
important research areas, like the search for extrasolar planets, the study of 
quasar hosts and, most relevant for this meeting, the surveying of 
nearby large scale structure in the Zone of Avoidance, in particular behind
the Galactic bulge.
\end{abstract}

\thispagestyle{plain}

\section{Introduction}
\citet{gonz98} developed and calibrated 
the Synthetic Field Method (SFM) to
derive the total extinction through disk galaxies. The method is based on the
number counts and colors of distant background field galaxies that can be seen
through the foreground object; 
it is the only method capable of determining
total extinction without {\it a priori} assumptions about the dust properties or its
spatial distribution. In principle, applying the SFM to the nearest galaxies,
like M~31 and the Magellanic Clouds, could offer a unique opportunity to study
the distribution of dust on small scales, to compare the distributions of warm
and cold dust, and to obtain the mass of molecular gas independently of the
highly uncertain CO to H$_2$ conversion factor. 
Unfortunately, background galaxies
cannot be easily identified through Local Group galaxies, even with the
resolution provided today by the Hubble Space Telescope 
\citep[{\sl HST};][]{gonz03}. This paucity of
detected background objects results in relatively large uncertainties on the
determination of the opacities (0.8 mags and  1.3 mags, respectively, in the
LMC and M~31; see Table 1).

In the case of M~31, each pixel in the WFPC2 images contains 50-100 stars, and
the background galaxies cannot be seen because of the strong surface brightness
fluctuations produced by nearly resolved stars. In the LMC, on the other hand,
there is only about one star every 6 linear pixels, and the lack of detectable
background galaxies is the effect of  a ``secondary" granularity, introduced
by structure in the wings of the point-spread function (PSF). As it often
happens, technological advances pose new problems, and going over the
atmosphere has had the unintended consequence that we now have to worry more
about the ``beauty" of the PSF.

The work discussed below about simulated observations of the LMC and M~31 
with diffraction-limited optical telescopes $\le$ 10-m has been 
published in an extended form before \citep{gonz03}. The predictions
for 20-m and 30-m telescopes, however, appear here in print for 
the first time. The section about the Galactic bulge is 
completely new. 

\section{M~31 and the LMC}

\subsection{Strategy}

To quantify the effects of the granularity, we produced artificial stellar
fields and measured their effect on the detectability of background galaxies.
These simulations were all produced in the $V$ passband, i.e., as if they had
been obtained through the {\sl HST} filters $F555W$ or $F606W$. The simulations had
several levels: first, we reproduced LMC and M~31 {\sl HST}
Wide Field Planetary Camera 2 (WFPC2) data, guided by
published luminosity functions and/or the star counts of the data, and
constrained by the mean and rms of the images. The second step was to simulate
the change in resolution offered, first by the smaller pixels of the Advanced
Camera for Surveys (ACS) on-board {\sl HST}, then by increasingly bigger, diffraction-limited
telescopes. This was achieved by changing the numbers of stars per pixel. Here,
we did two subclasses of simulations. In one case, we used Gaussian PSFs and
assumed that they were fully sampled. In the other, we used the Tiny Tim PSFs
for the two ACS cameras. Since the PSF for the ACS High Resolution Camera
(ACS-HR) is Nyquist-sampled in the visible, we used it to simulate the
observations with larger telescopes. Finally, we added the HDF-N to every
simulated stellar field, in order to assess the number of background galaxies
that we can expect to see in the absence of extinction. When few HDF-N galaxies
can be seen, it is impossible to know whether the paucity of real background
galaxies is due to crowding or extinction.

\begin{table}[!ht]
%\tablewidth{4.05in}
%\tablewidth{5.5in}
\caption{ \normalsize Error in opacity measurement (mags), 5.3 arcmin$^2$ FOV }

\medskip

\begin{tabular}{ccccc}
\tableline
\tableline
\noalign{\bigskip}
&\multicolumn{2}{c}{M~31 (2 hr exposure)}& \multicolumn{2}{c}{LMC (30 min exposure)}\\

\noalign{\smallskip}

Telescope size (m) & ``Real'' PSF & Gaussian PSF & ``Real'' PSF & Gaussian PSF\\ 
\noalign{\bigskip}

%\startdata
\tableline

\noalign{\bigskip}

2.4 (WFPC2) &1.3&&0.8&\\
\noalign{\smallskip}
0.8&&1.2&&0.7\\
\noalign{\smallskip}
1.1&&1.0&&0.6\\
\noalign{\smallskip}
2.4 (ACS-WFC) &0.9&&0.6&\\
\noalign{\smallskip}
1.6&&0.8&&0.6\\
\noalign{\smallskip}
2.3&&0.7&&0.4\\
\noalign{\smallskip}
2.4 (ACS-HRC)&0.9&&0.6&\\
\noalign{\smallskip}
3.&0.9&&0.7&\\
\noalign{\smallskip}
3.2&&0.6&&0.3\\
\noalign{\smallskip}
4.5&0.8&&0.6&\\
\noalign{\smallskip}
6.&0.8&&&\\
\noalign{\smallskip}
10.&0.8&0.5&0.5&0.2\\
\noalign{\smallskip}
20.&0.5&0.3&0.2&0.3\\
\noalign{\smallskip}
30.&0.5&0.3&0.3&0.3\\
\noalign{\bigskip}
\tableline
\end{tabular} 

\bigskip

\setlength{\lineskip}{5pt}
\renewcommand{\baselinestretch}{1.2}
\vspace*{0.3cm}\normalsize Note.  --- Except for {\sl HST} with WFPC2 and ACS-WFC, all telescopes are diffraction--limited. All Gaussian PSFs, and 
``real'' PSFs of telescopes 2.4--m and larger are well sampled. 
\label{taberr}
\end{table}
%\end{deluxetable}

%\end{document}

Figure 1 shows the galaxy counts in the M~31 simulations.
The field of view of 5.3 arcmin$^2$ is
equal to that of the combined three wide-field chips of the WFPC2. 
Changing the resolution of the telescope helps little, until the number of
stars per pixel falls below about one, where a jump can clearly be seen. This
jump corresponds to a change of regime when, with less than one star per pixel,
one encounters a situation similar to that of the LMC and secondary granularity
becomes strongly dominant. Indeed, the difference between Gaussian and
realistic PSFs turns significant only then.
Exposure time is never an issue with a realistic PSF; it only becomes important
with a Gaussian PSF and for telescopes 10-m and larger.

\begin{figure}[!ht]
%\plotone{m31.eps}
\plotfiddle{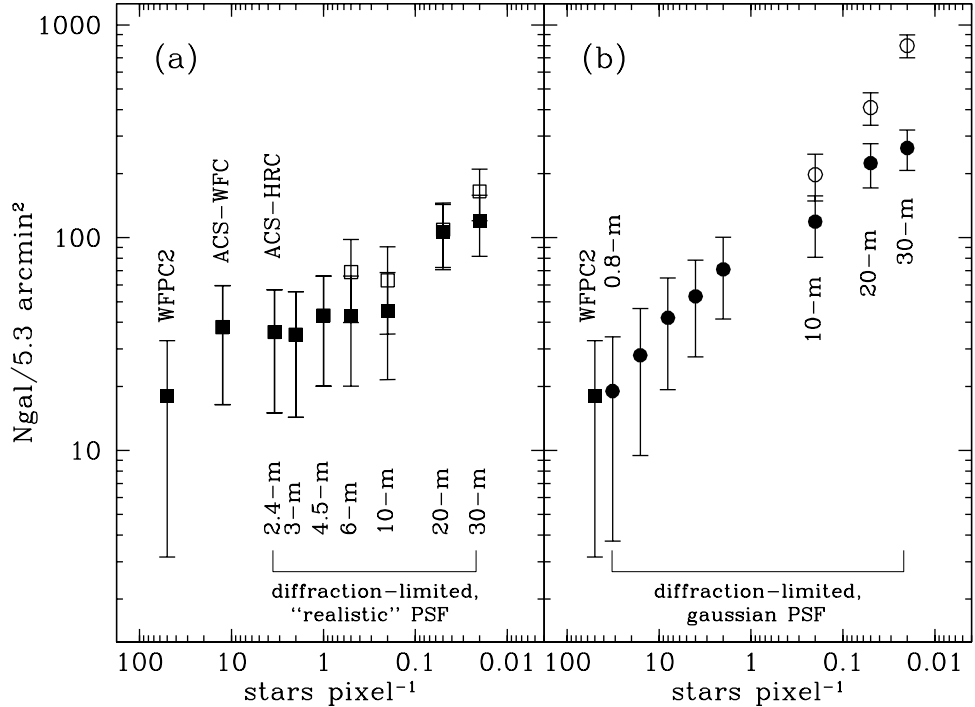}{10cm}{-90}{50}{50}{-195}{295}
\caption{Galaxy counts in M~31 simulations. (a):``Realistic" Tiny Tim PSF. 
(b): Gaussian, fully sampled PSF (except for
WFPC2 simulation). Open symbols: Noiseless data. Filled symbols: 2 hr exposures
with HST. This time should be scaled by $(2.4/D)^2$ for each pointing with
telescopes with different aperture $D$ in meters; however, since the
field-of-view will likely decrease also as $D^{-2}$, total exposure times should
stay roughly constant. Error bars are 3.5 times Poisson, 
in order to account for
field galaxy clustering.}
\end{figure}

\bigskip

Figure 2 displays galaxy counts in the LMC simulations. 
For ``small" telescopes, this is the case where one would  benefit most 
from both a Gaussian PSF and long exposure times. Very large telescopes, on the
other hand, would image about one star every 50 linear pixels (i.e., mostly
``empty" space). In those circumstances, PSF would be inconsequential, and 
exposure time would become an important factor.

\begin{figure}[ht!]
\plotfiddle{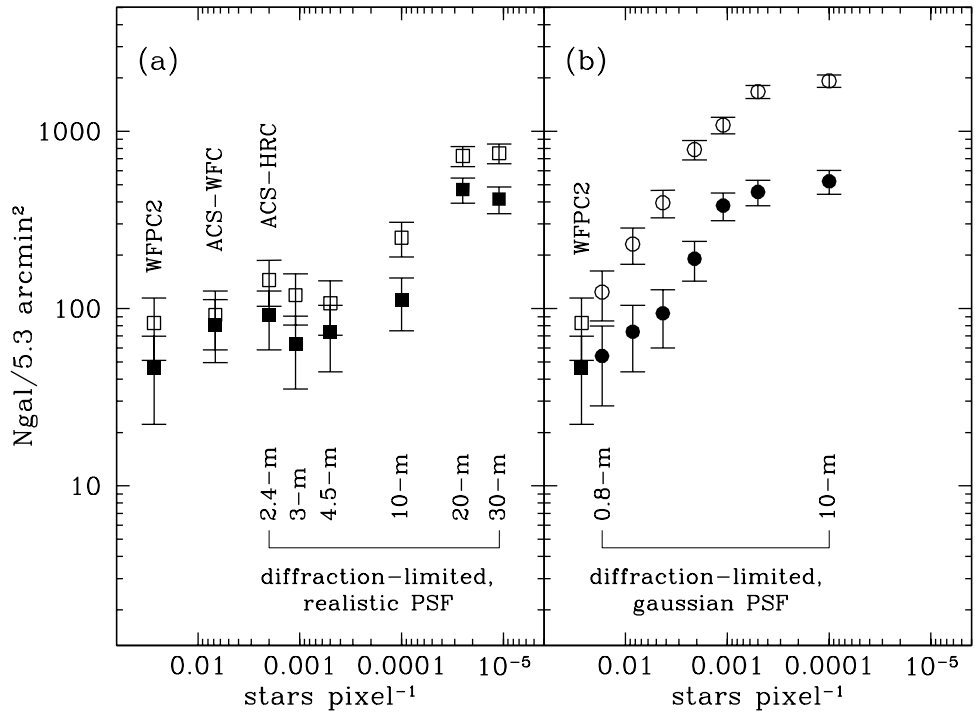}{10cm}{-90}{50}{50}{-195}{295}
\caption{Galaxy counts in the LMC simulations. 
The field of view of 5.3 arcmin$^2$ is
equal to that of the combined three wide-field chips of the WFPC2. (a):
``Realistic" Tiny Tim PSF. (b): Gaussian, fully sampled PSF (except for
WFPC2 simulation). Open symbols: Noiseless data. Filled symbols: 30 min
exposures with {\sl HST}; this time should be scaled by $(2.4/D)^2$ 
for telescopes with
different aperture $D$ in meters. Error bars are 3.5 times Poisson, in order to
account for galaxy clustering.}
\end{figure}

\subsection{The model} \label{mod}

The behavior of background galaxy detection as a function of 
distance to the foreground galaxy and telescope size 
can be understood 
in a simple statistical way, especially in the case of a Gaussian
PSF \citep[for details, see][]{gonz03}. Assuming that 
all foreground stars are identical, and have an 
intrinsic flux $f_*$ (in
photons sec$^{-1}$ received at the detector), a background galaxy
will be detected with a
signal-to-noise ratio equal to the
quotient of the number of photons received from that background object
($f_{bg}t$) divided by the noise produced by Poisson fluctuations in
both the stellar photons and the number $n$ of foreground stars:

\begin{equation}
\left(\frac{S}{N}\right)_o~=~\frac{f_{bg}t}{t\sqrt{nf_*^2+n^2f_*}}~=~\frac{f_{bg}}{\sqrt{nf_*^2+n^2f_*}}
\end{equation}

When telescope resolution is improved and pixels are made 
$L$ times smaller (on a side), signal-to-noise ratio 
becomes:

\begin{equation}
\left(\frac{S}{N}\right)_n~=~\frac{Lf_{bg}}{\sqrt{L^2nf_*^2+n^2f_*}}
\end{equation}

If the same foreground stellar population is pushed $d$ times
farther away, the signal-to-noise ratio results in:

\begin{equation}
\left(\frac{S}{N}\right)_d~=~\frac{Lf_{bg}}{\sqrt{\frac{L^2nf_*^2}{d^2}+n^2f_*d^2}}.
\end{equation}

\begin{figure}[ht!]
\plotfiddle{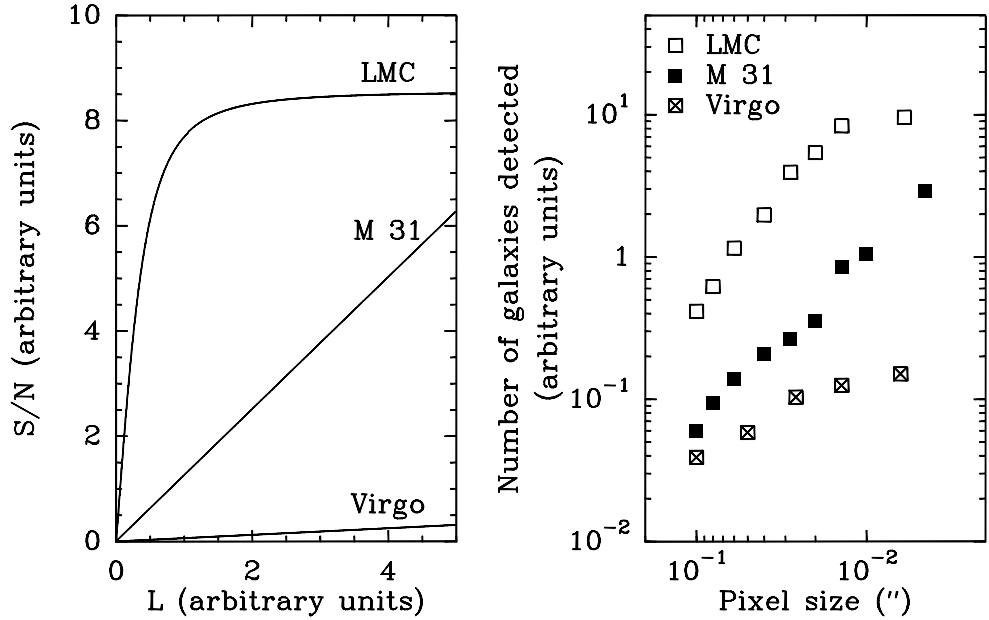}{8.5cm}{-90}{60}{60}{-200}{305}
\caption{Comparison of the behavior of eq. (3) ({\it left}) with the number of 
background galaxies detected through the LMC, M~31, and Virgo 
({\it right}); an arbitrary constant has been added to the logarithm of the
number of background galaxies seen through each object to facilitate
the comparison of the data with the model.}
\end{figure}

\bigskip

The behavior of this expression as a function of pixel size (set by $L$) for
the same stellar population at three different distances $d$ is shown on the
left panel of Figure 3. The value of $f_*$, the average stellar flux, 
and $n$, the number of stars per pixel, have been chosen to 
reproduce (although with arbitrary units) the conditions in 
the LMC; we further mimic the conditions found in 
M~31 and Virgo by changing only $d$, the relative distance.  
The qualitative agreement between the left and right panels
of Figure 3 is
remarkable, especially given the extremely simple model we
have adopted for the noise.

\section{Application to the Galactic bulge}

The Galactic bulge occupies approximately one thirty-seventh of the 
sky, or the area of more than 5000 Suns! It constitutes almost one-third of 
the Zone of Avoidance. Although quantifying the extinction 
suffered by galaxies seen
through the bulge has interest, it is potentially much more
important to characterize the distribution of nearby galaxies 
behind the bulge. 
Except at very low latitudes ($b\la 2\deg$), 
the main obstacle to finding galaxies in this region of the sky is
stellar crowding, more than dust extinction
%\citep[][Kraan-Korteweg 2004, personal communication]{kraa00}.
(Kraan-Korteweg \& Lahav 2000; Kraan-Korteweg 2004, personal communication).
Going to the near-IR to try to circunvent extinction only makes
stellar crowding worse, since the old and metal rich population 
of the bulge emits strongly in this spectral region.

Short of doing simulations to try to quantify background galaxy
detection rates behind the bulge in the near-IR, we have performed
two little exercises. First, we notice (Tully 2004, personal communication) 
that the galaxy count-rate per interval of sin $\mid b \mid$ in the  
Two Micron All-Sky Survey (2MASS)
is flat up to $\mid b \mid \sim 13\deg$ and then 
decreses steadily; 
%until about 20 times fewer 
%are found at $b \sim 1\deg$; 
almost no galaxies are seen at latitudes lower than $\mid b \mid \sim 1\deg$.
On the other hand, in the $K$-band, the 
surface brightness of the bulge is 100 times larger  
at $\mid b \mid \sim 1\deg$ than at 13$\deg$ \citep{kent91}.
This means that, in principle and in the absence of
extinction, with a linear resolution 10 times
better than that of 2MASS --which would allow us to image
one hundred times fewer foreground bulge stars per
pixel--, it would be possible to see  
as many background galaxies at $b \sim 1\deg$ as can be 
detected at 13$\deg$. 

The pixels of the released images of 2MASS are
1$\arcsec$, and we have measured an average FWHM of $\sim 3\arcsec$.
The resolution of the {\sl HST} NICMOS Camera 2 (NIC2), 
with a pixel scale of 0\farcs075 
pixel$^{-1}$, is already in the right ballpark; the problem
is that the NIC2 field-of-view is very small ($19\arcsec \times 19\arcsec$) 
and, of course, that the PSF is not Gaussian 
\citep[see, for example,][]{kris98}. 

The other exercise consists in the application of our toy model,
only this time changing the ``stellar population,'' that is, the 
brightness of the average star and the stellar density. 
%as well as
%the distance. We use one sixth of the distance to the LMC.
Stellar population synthesis models \citep{bruz03} show that 
1 M$_\odot$ of a population with solar metallicity and 10$^{10}$ yr (i.e., like 
that of the bulge) has an absolute $K$ mag $M_K \sim 3.7$. 
%Conversely, for the LMC population that we have used as a template for our 
%little model in \S \ref{mod}, \citet{elso97} have determined an age of 
%about 2 Gyr and $Z = 0.004$. According to \citet{bruz03}, 1 M$_\odot$
%of such a population has $M_K \sim 2.7$ and ($V! - !K$) $\sim$ 2.2.
On the other hand, the LMC population that we have used as a template for our
little model in \S \ref{mod} 
%has a surface brightness at the 
%visual filter $F606W\ 
%\mu_{F606W} = 21.7$ \citep{gonz03}, 
has an age of
about 2 Gyr and $Z = 0.004$ \citep{elso97}. According to \citet{bruz03}, 
%such a population 
%has a ($V - K$) color $\sim$ 2.2, which implies that
%$\mu_K \sim$ 19.5; also after \citeauthor{bruz03}, 
1 M$_\odot$ of this 
population has $M_K \sim 2.7$, i.e., it is 1 mag brighter at $K$
than the bulge population. 
The template LMC population has a surface brightness 
$\mu_K \sim$ 19.5, while the 
Milky Way bulge has 
$\mu_K \sim$ 15 at $\mid b \mid \sim 0-1\deg$, and $\sim$ 15.5 at 
$\mid b \mid \sim 4\deg$. For our toy model, what all this means 
is that the 
old and metal rich population, which is intrinsically 1 mag fainter per solar 
mass than that of the LMC disk, produces a bulge surface brightness 
that is between 4 and 4.5 mag brighter. Consequently, the stellar 
density of the bulge must be 63 times and 158 times 
higher than that of the LMC disk at, respectively, $\mid b \mid \sim 4\deg$ and 
$\mid b \mid \sim 0-1\deg$. Figure 4 shows the change of
background galaxy detection rate
with resolution (again with units that are arbitrary, and 
different from those used in Figure 3) 
for the bulge stellar population, at these 
two different Galactic latitudes 
and in the near-IR. 
The population is set at one-sixth the distance to the LMC; 
it is assumed to be the 
same at both latitudes, except for the stellar 
density, and it is characterized by an average 
brightness of a single star that is 40\% of the one used 
in \S \ref{mod}. 
Predictably,
the curve for $\mid b \mid \sim 4\deg$ not only
reaches higher, but also --although it is hard to
notice-- rises faster.
The extremely fast upsurge of 
the detection rate in both cases, however, confirms 
that a telescope in space as 
small as the Hubble 
could do the job, provided it had adequate optics and a large
field of view. Also,  
by comparison with the simulations of the LMC and
M~31, we could expect 
satisfactory results with a ``realistic," non-optimized, PSF already with 
a 10-m class telescope in space.

\begin{figure}[ht!]
\plotfiddle{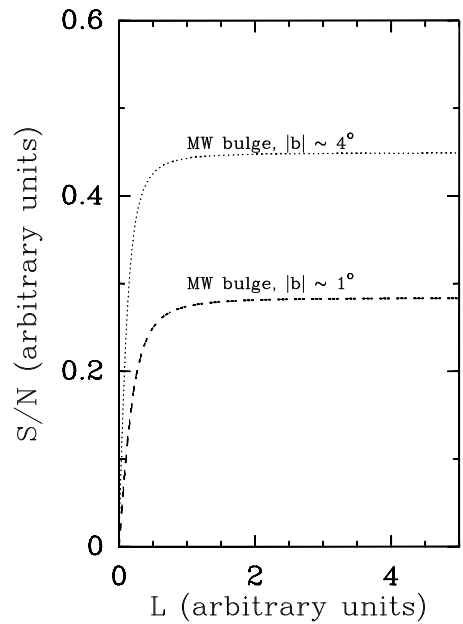}{8.5cm}{-90}{60}{60}{-110}{305}
\caption{Behavior of eq. (3) for bulge parameters. 
{\it Dashed line:} $\mid b \mid \sim 1\deg$; 
{\it dotted line:} $\mid b \mid \sim 4\deg$.
The distance of the foreground population is 1/6 of that to the LMC.
To mimic an old and metal-rich stellar population, as well as observations
in the near-IR, the average flux of an individual star is taken to be 
40\% of the one used in Figure 3. Always comparing with the
LMC disk, the stellar density is, respectively, 158 and 63 times higher. 
} 
\end{figure}

\section{Conclusions} 
We have investigated how large and very large, 
diffraction-limited, optical and infrared telescopes
in space would improve the detection of background galaxies behind Local
Group objects, including the Galactic bulge. 
For example, with a 20-m optical telescope, the total extinction through the LMC
could be determined with an error of 0.2 mags in a field of view of 5.3
arcmin$^2$, in about 30 min total exposure time, 
independently of the quality of
the PSF. Likewise, the number of galaxies seen behind the
Galactic bulge should increase significantly, in spite of the
high extinction in the midplane, with a 10-m infrared telescope; 
important strides could be made behind the bulge 
even with a smaller telescope,
provided the PSF were close to Gaussian.
Indeed, probably 
more important than a large aperture, a well-behaved, 
well-characterized PSF would facilitate in general the
detection of faint objects in crowded fields, and greatly benefit several
important research areas, 
like the determination of total extinction in Local Group objects, the search for extrasolar planets, the study of 
quasar hosts and, most relevant for this meeting, the surveying of 
nearby large scale structure in the Zone of Avoidance, in particular behind
the Galactic bulge.

\end{document}